\documentclass[aps,showpacs]{revtex4}
\usepackage{epsfig}

\begin{document}
\title{ Entropy and multifractal analysis of multiplicity distributions from
$pp$ simulated events up to  LHC energies}

\author{M. K. Suleymanov}
\altaffiliation{on leave of absence from LHE JINR, Dubna,Russia}
\email{mais@sunhe.jinr.ru}
\affiliation{Nuclear Physics Institute, Academy of Sciences of the Czech Republic,
\v{R}e\v{z}, Czech Republic}

\author{M. \v{S}umbera}
\email{sumbera@ujf.cas.cz}
\affiliation{Nuclear Physics Institute, Academy of Sciences of the Czech Republic,
\v{R}e\v{z}, Czech Republic}

\author{I. Zborovsk\'{y}}
\email{zborovsky@ujf.cas.cz}
\affiliation{Nuclear Physics Institute, Academy of Sciences of the Czech Republic,
\v{R}e\v{z}, Czech Republic}

\begin{abstract}
Using three different Monte Carlo generators of high energy
proton-proton collisions (HIJING, NEXUS, and
PSM) we study the energy dependence of multiplicity distributions of charged
particles including the LHC energy range.
Results are used for calculation of the information entropy,
Renyi's dimensions and other multifractal characteristics of particle production.
\end{abstract}


\maketitle

\section{Introduction}

Energy dependence of multiplicity distribution (MD) of particles
produced in high energy collisions of hadrons and nuclei is
important issue of multiparticle dynamics \cite{Dremin_PR}. Though
MD contains information about particle correlations in an
integrated form it provides general and sensitive tool to probe
the dynamics of the interaction. Analysis of the multiplicity data
from $pp$ and $AA$ collisions will be thus important part of the
physics programme
at the Large Hadron Collider (LHC) at CERN. In this new energy domain some
basic questions concerning MD remain still open.
Current theoretical understanding of the interplay between soft
and semihard mechanisms of particle production
is insufficient to provide reliable estimates even of the
elementary quantity characterizing probability distribution ${P_n}$ of
the $n$-particle event - the average multiplicity of particles
$<$$n$$>$ produced in $pp$ collisions. Different scenarios vary
over a~wide range of values \cite{GU99}.
Notwithstanding such uncertainty, large multiplicity of particles
makes it feasible to study specific characteristics of MD which are relevant
to fractal properties of the multiparticle dynamics. Fractality is
usually connected  with investigations of multiplicity in limited
intervals of phase-space and demonstrated as a power-law in resolution dependence of
the multiplicity moments \cite{Hwa1,Sarcevic}. Such behaviour is characteristic for effects of
nonstatistical fluctuations \cite{Miyamura} and/or intermittency \cite{Bialas}.
Besides fractality, there exists evidence in favour
of multifractality  \cite{Peshanski,Hwa2} in the structure of inelastic events.
Some authors brought suggestions \cite{DeWolf} that the property could
be expressed in simple thermodynamic terms.
Namely, it was shown \cite{Bershadskii} that constant specific heat, widely used
in standard thermodynamics,
reflects multifractal character of various stochastic systems in a reasonable approximation.
In hadronic interactions, its constancy relative to the q-order number of the multifractal
characteristics (the generalized dimensions) was indicated  in Ref. \cite{Parashar}.
Universality of the multifractal specific heat with respect to various hadron-nucleus
reactions
was demonstrated \cite{Ghosh} by means of the method proposed by Takagi \cite{Takagi}.
Though exploiting similar variables and methods supports multifractal interpretation of data,
conclusion about the methodology is far from being unique.
In such situation, alternative approaches to the above aspects of
multiparticle production are also needed. They can be helpful in better understanding of
multifractality in high energy collisions and useful in
extracting dynamical origin of its
basic phenomenological observation - the power law dependence of the corresponding
measures with respect to the fractal resolution.

In this paper we employ and further extend our earlier study of MD
\cite{SSZ,PSSZ} based on robust characteristics of MD such as
Shannon's {\it information} entropy
\begin{equation}\label{entropy}
S=-\sum P_n \ln  P_n
\end{equation}
and its generalization, R\'enyi's order-q {\it
information } entropy \cite{Renyi,Feder}
\begin{equation}\label{ientropy}
I_q=\frac{1}{1-q}\ln\sum (P_n)^q
\end{equation}
for which $I_1=S$.
They contain information on the multiplicity moments
$<$$n^k$$> = \sum {n^kP_n}$ in a non-trivial way.
The outline of the paper is as follows. In Section 2 we show how
regularities in energy dependence of the Shannon entropy provide
independent constraint on the energy dependence of $<$$n$$>$.
Section 3 is devoted to multifractal interpretation of observed
regularities in terms of the R\'enyi's order-q  entropy.
We use three different Monte Carlo (MC) generators of high energy
proton-proton collisions, HIJING \cite{HIJING}, NEXUS \cite{NEXUS,Werner_QM01},
and PSM \cite{PSM} to study  energy dependence of MD up to the
LHC energies.
Results of MC simulations are reported in Section 4.
Summary and discussion of the results is presented in Section 5.

\section{Entropy}
Entropy is important characteristic of systems with many degrees
of freedom. It seems quite natural to use it in description of
high energy multiparticle production processes. In particular,
entropy of MD is an effective variable characterizing inelastic
collisions with many particles produced. The simple relation \cite{SSZ}
\begin{equation}\label{S_lnN}
S~- \ln<n>~ \simeq -\int_0^{\infty}\psi(z)\ln\psi(z)dz
\end{equation}
between the entropy $S$ and the average multiplicity $<$$n$$>$ is
valid with good accuracy for large enough $<$$n$$>$.
The function $\psi(z)$ is related to the probability $P_n$ by the formula
\begin{equation}\label{KNO1}
\psi\left(z=\frac{n}{<n>}\right) \equiv <n>P_n.
\end{equation}
The KNO scaling \cite{KNO} of MD
postulates energy independence of $\psi(z)$
what can be expressed according to Eq. (\ref{S_lnN}) as follows
\begin{equation}\label{KNO}
S~- \ln<n> ~ \simeq ~ const.(\sqrt s).
\end{equation}
Last relation implies that all {\it information}
about the energy dependence of MD is contained in its first moment $<$$n$$>$.
Such behaviour was indeed observed in hadron-hadron collisions
up to the ISR energy range \cite{ISR}. However, beyond
this range, the KNO scaling of MD of charged hadrons produced
in $pp/p\bar{p}$ collisions was shown to be significantly violated
\cite{UA5a,UA5b}. The observed scaling violation, i.e. break down of
Eq. (\ref{KNO}) results via Eq. (\ref{S_lnN}) in non-trivial
correlation between $S$ and $<$$n$$>$.
Experimental situation concerning the difference $S-\ln<\!n\!>$
calculated for MD of charged particles in the full phase space is illustrated in Fig.~1.
Here, due to the charge conservation, $<\!n\!>=<\!n_{ch}/2\!>$ is half of the
average multiplicity of charged particles.
Recently published Tevatron
data \cite{E735} from $p\bar p$ collisions are also included in the figure.
The values of the difference (\ref{S_lnN}) show clear increase with energy and
thus confirm violation of the  KNO scaling observed at lower ISR energies.

Let us note that increase of $S-\ln<\!n\!>$ can not continue ad
infinitum. It was shown \cite{SSZ} that this difference is
asymptotically bounded by unity from above. The extremal value of the r.h.s. of the
relation
(\ref{S_lnN}) is obtained from the principle of maximal entropy
applied to all continuous functions $\psi(z)$. If the functions (\ref{KNO1})
satisfy the conditions
\begin{equation}\label{KNO2}
\int_0^{\infty}\psi(z)dz = \int_0^{\infty}z\psi(z)dz = 1
\end{equation}
valid for all normalized MD with
given $ <\!n\!>$, the extremum is reached for the KNO function
$\psi(z)=\exp(-z)$ which corresponds to the geometrical
distribution. The difference
\begin{equation}\label{bound}
S-\ln<\!n\!>   \le  (1+<\!n\!>)\ln(1+\frac{1}{<\!n\!>}) = 1+
O(<\!n\!>^{-1})
\end{equation}
is thus bounded from above by the expression
which is calculated for the geometrical MD. The bounding values are represented by the
curve in Fig.1. Last relation shows that energy dependence of the {\it
information} entropy can be used as an independent constraint on
energy dependence of the average multiplicity.

Energy dependence of the entropy $S$ in high energy collisions was first
studied in Ref. \cite{SSZ}. Using MD of charged secondaries
produced in $pp$ and $p \bar p$ collisions in the energy range
$\sqrt{s}\le 900$~GeV, monotonous increase of $S$ with
center-of-mass energy $\sqrt{s}$ was found.
For $\sqrt{s} > 20$~GeV, the linearity
\begin{equation}\label{iregularity}
S= D_1 Y_m
\end{equation}
with the maximum rapidity  $ Y_m=\ln(\sqrt{s}/m_{\pi})$
of the hadrons produced is valid.
Here $m_{\pi}$ is the pion mass and $D_1$ is an energy independent
constant.
Recently published Tevatron data \cite{E735} extend the validity
of these findings up to $\sqrt{s}=1.8$~TeV. The experimental
situation is shown in Fig.~2 where the published errors of MD have
been taken into account. Value of $D_1$ is within $\pm2\%$ error
band consistent with the predicted one $D_1\equiv S/Y_m = 0.417\pm
0.009$ \cite{SSZ} shown by the full/dashed lines in Fig.~2. On
closer inspection, however, one can see that $D_1$ calculated form
E735 data is systematically above the UA5 values. It is connected
with higher tails of the E735 MD as compared to the UA5 data. The
discrepancy might be due to the sizable systematic uncertainty in
both E735 and UA5 since in both experiments the full phase space
MD were obtained by computer simulation from data measured in a
restricted range of rapidity.

As the bounding value in (\ref{bound}) tends to unity, the
observed monotonous increase of the entropy $S=D_1
\ln(\sqrt{s}/m_{\pi})$, if valid at higher energies, will also
govern the energy dependence of $<$$n$$>$:
\begin{equation}\label{Mult}
<n> \approx \exp(S)= (\sqrt{s}/m_{\pi})^{D_1} .
\end{equation}
Such asymptotic power law behaviour of the average multiplicity
should be contrasted with other approaches to multiparticle
production. In particular in Ref. \cite{GU99}, $<$$n$$>$ was
predicted to increase as a~second order polynomial in $\ln s$.
Difference between these two predictions
will be substantial at the top energy of LHC, because according to
Eq. (\ref{Mult}) $<$$n$$>$$\approx 110$ at $\sqrt{s}=14$~TeV, while according to the
parametrization used in Ref. \cite{GU99} $<$$n$$>$$\approx70$ at this energy.

\section{Multifractality}

Fractal geometry is nowadays widely used in many branches of
physics \cite{Feder}. In astronomy analysis of galaxy and cluster
distributions it has led to surprising result that galaxy
correlations up to $150h^{-1}Mpc$ are scale invariant and not
homogenous \cite{Pietronero}. In multiparticle dynamics methods
introduced originally for description of the fractal properties of
stochastic systems \cite{Paladin} are used extensively
\cite{Dremin_PR}. In particular, study of MD in small rapidity bins
using the scaled factorial moments $F_q$ revealed typical linear behaviour
\begin{equation}\label{intermitt}
\ln F_q= -\tau_q \ln\delta
\end{equation}
in terms of the bin resolution $\delta$. Such behaviour was interpreted as
(multi)fractal property of particle production.
Another approach to the multifractality is directly connected to the R\'enyi's order-q {\it
information } entropy $I_q$ (\ref{ientropy}).
Energy dependence of $I_q$ in high energy collisions was first studied in Ref. \cite{PSSZ}.
Using MD of charged secondaries
produced in $pp$ and $p \bar p$ collisions in the energy range
$\sqrt{s}\le 900$~GeV it was found that increase of $I_q$ with
center-of-mass energy $\sqrt{s}$ is similar for various values of
$q$.
For $\sqrt{s} > 20$~GeV, the observed asymptotic linearity
\begin{equation}\label{iregularity2}
I_q= D_q Y_m
\end{equation}
with maximum rapidity of the hadrons produced, $ Y_m=\ln(\sqrt{s}/m_{\pi})$,
generalize Eq. (\ref{iregularity}).
Let us briefly summarize multifractal interpretation of
Eq. (\ref{iregularity2}).
In contrast to standard intermittency and
multifractal analysis we consider fractal resolution $\delta$ related to the total
energy $\sqrt{s}$ available and not to the phase-space bining.
During hadron-hadron interaction with many particles produced, the energy
dissipates into $N=\sqrt{s}/m_{\pi}$ discrete sites each of the size $m_{\pi}$.
The site labeled by $n$ is occupied with the probability $P_n$.
Since most of the sites are unoccupied the overlay of many
inelastic events can be visualized as a fractal with overall extent
$\sqrt{s}$ characterized  by  a local mass
distribution function. The sufficient condition to produce
such self-similar (hierarchical) structure is that the probability
$P_n$ exhibits some type of scale invariant behaviour. In particular,
if at sufficiently high energies, i.e. at high enough resolution
$\delta=1/N$, the $P_n$ acquires a power law dependence on
the resolution $\delta$, the quantity $\sum (P_n)^q$ will scale
with $\delta$ like
\begin{equation}\label{fresolution}
\sum (P_n)^q \sim \delta^{-(1-q)D}.
\end{equation}
If  $D>0$ is independent of $\delta$, one usually speaks about fractality  \cite{Feder}
of the distribution $P_n$. The multifractals
\cite{Feder,Paladin} generalize the notion of fractals for $D=D_q$
depending on $q$. Spectrum of generalized dimensions $D_q$ which
for multifractals is a decreasing function of $q$ \cite{Paladin}
has the following meaning. $D_0$ corresponds to the capacity (box
dimension) of the support of the measure $P_n$, {\it information
dimension} $D_1$ \cite{Renyi} characterizes scaling of the
information entropy (\ref{entropy}) and $D_q$'s for integer $q
\geq 2$ can be related to the scaling behaviour of $q$-point
correlation integrals \cite{Feder,Paladin}.
The observed  approximate independence of the generalized dimensions
\begin{equation}\label{Dq}
D_q= -I_q/\ln\delta
\end{equation}
on the energy (resolution $\delta$) as well as  their decrease with increasing $q$
show \cite{PSSZ} that for $\sqrt{s}\ge 20$~GeV the full phase space MDs of
charged particles from non-single-diffractive hadron-hadron
collisions are indeed multifractal.
Let us note that multifractality besides predicting $D_q$ to be
decreasing functions of $q$ does not, in general, provide any
further information about the $q$-dependence of the spectrum of
the generalized dimensions $D_q$. In particular knowledge of say $D_1$
and $D_2$ is insufficient to predict scaling behaviour of the
higher $q$ correlation integrals. It is thus gratifying to find
out that this could be, at least in principle, achieved within
interpretation of multifractality in thermodynamical terms
\cite{Bershadskii}. Latter is based on analogy between l.h.s. of
Eq. (\ref{fresolution}) and partition function
\begin{equation}\label{part_fun}
Z(q)\equiv\sum (P_n)^q
\end{equation}
with $q$ playing the r\^{o}le of inverse temperature $q \equiv
T^{-1}$ and $V \equiv -\ln\delta$ representing volume. The
thermodynamic limit of infinite volume $V \rightarrow \infty$ is
then equivalent to the limit of increasing resolution $\delta
\rightarrow 0$. In the constant specific heat approximation the q-dependence of the
generalized dimensions $D_q$ acquires particularly simple form \cite{Bershadskii}
\begin{equation}\label{Berash}
D_q \simeq (a -c) + c\frac{\ln q}{(q-1)}.
\end{equation}
The coefficient $c$ represents multifractal specific heat and
$a = D_1$.
Regular behaviour of this type is
expected to occur for multifractals for which, in classical analogy with specific
heat of gases and solids, the multifractal specific heat $c$
is independent of temperature \cite{Landau} in a wide range of $q$.

We have examined validity of the approximation given by
Eq. (\ref{Berash}) for $D_q$ defined by Eqs. (\ref{iregularity2}) and
(\ref{ientropy}). In Fig.~3a we present q-dependence of generalized
dimensions calculated from the Tevatron data \cite{E735} at
$\sqrt{s}=300$, 546, 1000, and 1800~GeV.
One can see from the
figure that the values of $D_q$ reveal indeed linear increase as a
function of $\ln(q)/(q-1)$. This behaviour makes it possible to
define the slope parameter $c$ in the region $q\ge 1$. Similar
$D_q$ dependencies for data from CERN ISR and $Sp\bar{p}S$
Collider experiments \cite{ISR,UA5a,UA5b} are shown in Fig.~3b.
The dashed line coincides with the full line in Fig.~3a indicating
position of $D_q$ values calculated from E735 data. Both
data sets obtained by Tevatron  and CERN experiments reveal
approximately the same slope while their intercepts are mutually
shifted. The shift is due to larger values of $D_1$ for E735 data,
as already shown in Fig.~2. As pointed in the previous section, the
discrepancy in intercepts is
connected with systematically larger high multiplicity tails of
the data from Tevatron when compared to the data from the CERN
$Sp\bar{p}S$ Collider. This might be connected with the mutual
systematic uncertainties of the experimental procedures when
extending measured data into the full phase space region.

Fitting the slope parameters for $q\ge 1$ at each separate energy,
we have determined the values of multifractal specific heat $c$.
The results are presented in the lower part of Fig.~2. The errors
in determination of $c$ were calculated from error bars of MD
quoted in literature. They represent $\sim 10-15\%$ of the
established values. For $\sqrt{s}\ge 20$~GeV, the multifractal
specific heat is within the estimated errors approximately energy
independent quantity and achieves the value $c\approx0.08$. Note
that this number practically coincides with the slope obtained
from the electron-positron multiplicity data \cite{LEP} what is
indicated by the full line in Fig.~3b.
Since our study concerns the full phase
space and it is performed in a different sense than the usual
intermittency analysis, we obtain smaller value of $c$ in
comparison with the specific heat ($c\sim 0.26$) determined from
multifractal properties of the factorial moments \cite{Bautin}.
While the multifractal specific heat reported in Ref. \cite{Ghosh} reveals
some kind of universality with respect to various interactions, energy dependence of
$D_q$ obtained with the same method \cite{Takagi} seems to be significant.
Contrary to this, our method gives smaller values of the generalized dimensions
$D_q$ which are approximately energy independent for $\sqrt{s}\ge 20$~GeV.

Energy independence of the multifractal specific heat
confirms that, in addition to the information dimension $D_1$,
there appears to be yet another parameter $c$ which could be used
as universal characteristic of particle production in
hadron-hadron interactions at high energies. Knowing $D_1$ and $c$,
all other $D_q$ can be thus deduced from Eq. \ref{Berash}.

\section {Monte Carlo simulations}

We have exploited three different Monte Carlo models
in our investigation of the proton-proton interactions beyond up
to date accessible energies.

1.  The HIJING~\cite{HIJING} Monte Carlo code is based on
QCD-inspired models for multiple jets production. It allows to
study jets and the associated particle multiplicities. The model
includes minijet production, soft excitation, nuclear shadowing of
parton distribution functions and jet interaction in the dense
nuclear matter.

2.  The NEXUS~\cite{NEXUS,Werner_QM01} Monte Carlo code is based
upon the hypothesis that the behaviour of the high energy
interactions is universal. Basic building blocks of hadron-hadron
or nucleus-nucleus scattering are parton ladders coupled softly to
the nucleons. It relays on a~consistent multiple scattering
approach in the sense that most of the dynamics follows from
a~formula for the total cross section expressed in terms of cut
diagrams.

3. The Parton String Model (PSM)~\cite{PSM} includes in its
initial stage both soft and semihard components which lead to the
formation of color strings. Collectivity is taken into account
considering the possibility of strings in color representations
higher than triplet or anti-triplet by means of string fusion.
String breaking leads to the production of secondaries.

We have analyzed $\sqrt{s}$-dependence of the
average multiplicity $<$$n_{ch}$$>$, the entropy $S$, the
difference $S-\ln(<$$n_{ch}$$>\!/2)$, and the multifractal specific heat $c$
for the charged particles in the energy interval 25~GeV$<\sqrt{s}<14$~TeV.
The quantities have been obtained for 14 points from this interval using 10000
simulated events for each one.
The results of the energy dependence of $<$$n_{ch}$$>$ are shown
in Fig.~4. Though all three models give similar values in the range
$\sqrt{s}<1$~TeV, their predictions significantly differ for $\sqrt{s}>$1~TeV.
The average multiplicity simulated by HIJING
and PSM codes is well approximated by the power dependence
\begin{equation}\label{PSM_mult}
<n_{ch}> =as^b
\end{equation}
with $b\simeq 0.018$ for HIJING and  PSM. The NEXUS predictions fall well bellow
and can be fitted much better by the second order polynomial in logarithm
$\sqrt{s}$:
\begin{equation}\label{NEXUS_mult}
<n_{ch}>= a_0+a_1\ln\sqrt{s} + a_2\ln^2\sqrt{s}.
\end{equation}
The above parametrization corresponds to the prediction of the
average multiplicity given in Ref. \cite{GU99}.

Let us now compare both parametrizations from the point of view of
the observed regularity in the information dimension $D_1$. The energy
dependence of $D_1=S/Y_m$ is displayed in Fig.~5. For HIJING and
PSM the behaviour is consistent with the experimentally observed
ratio $S/Y_m=0.417\pm0.009$ extrapolated towards super-high energies.
This supports the conjecture that ${\it
information}$ entropy per unit rapidity should stay constant in
the full phase space up to the LHC energy region and even beyond.
In such case, energy dependence of the average multiplicity should
achieve the asymptotic power law behaviour (\ref{Mult}). The
HIJING and PSM simulations of the average multiplicity
(\ref{PSM_mult}) confirm this tendency. On the other hand, results
of simulations with the NEXUS code show slow but continuous
decrease of $D_1=S/Y_m$ with the energy. The decrease of $D_1$ is
connected with non-power like, namely the logarithmic energy
dependence (\ref{NEXUS_mult}) of the average multiplicity. Using
the general limit (\ref{bound}) on MD one can say: either the
entropy $S$ must slow down and violate the entropy scaling
(\ref{iregularity}) as indicated by the NEXUS model, or, providing the
entropy scaling stays valid, the average multiplicity
must grow faster, similar as in the HIJING and PSM models, and
violate thus the parametrization (\ref{NEXUS_mult}) of
Ref. \cite{GU99}.

The above statement relies in details on the energy dependence of
the difference $S-\ln(<n_{ch}>\!/2)$ which is depicted in Fig.~6.
Though for all energies considered the difference does not reach
the bounding value (\ref{bound}), its saturation in the LHC energy
range is almost complete for the PSM model.
The NEXUS prediction falls well bellow at the LHC energy and  is closer to the
KNO-scaling behaviour given by Eq. (\ref{KNO}). In the case of the
HIJING we have checked the sensitivity of $S-\ln(<n_{ch}>\!/2)$ to the
interactions dynamics, changing in particular number of the
produced jets $N_{jets}$. There are points of regime changes
clearly visible in Fig.~6. They correspond to various maximal
number of produced jets, which had been set to $N_{jets}=$0,1, and
5 for $\sqrt{s}<0.1$~TeV, 0.1~TeV$<\sqrt{s}<1$~TeV,  and
$\sqrt{s}>1$~TeV, respectively. Character of these changes
demonstrates that the quantity $S - \ln(<$$n_{ch}$$>\!/2)$ is sensitive to
the dynamic of the interaction, in particular, to the number of
produced jets. Increase of $N_{jets}$ results in larger number of
parton-parton collisions which play role in the KNO scaling
violation \cite{E735}. The violation seems however not to destroy
the self-similarity of parton dynamics as manifested by the
regularity in behaviour of the {\it information } entropy of MD
including the data from Tevatron (Fig.~2).

Multifractal character of particle production consistent with approximate energy
independence of the multifractal specific heat $c$
is predicted by the PSM, HIJING,  and NEXUS models in the LHC energy
region and beyond. We demonstrate this in Fig.~7.
The tendency towards a constant value of $c$ confirmed by the simulation
results is  similar with the experimental situation shown in
Fig.~2. It will be interesting to see whether this will
really hold true for the forthcoming data from LHC.

\vspace{1cm}

\section{Summary and discussion}

Having enumerated experimentally observed features of MD in
multi-hadron production such as regularities in the {\it
information} entropy and  generalized dimensions, we have studied
the multifractality of inelastic hadron-hadron collisions at high
energies. We have applied the multifractal specific heat approach
to characterize this property in a quantitative way. The
generalized dimensions were calculated using the resolution
$\delta=m_{\pi}/\sqrt{s}$ defined as minimal fraction of the
center-of-mass energy which can be carried away by the outgoing
hadrons. The experimental data on MD are consistent with the
constant specific heat approximation $c\sim 0.08$ with respect to
both, the multifractal temperature $T\equiv q^{-1}$
\cite{Bershadskii} and the energy $\sqrt{s}$.

We have extended our study of MD in $pp$ collisions up to LHC
energies exploiting three different Monte Carlo generators of high
energy hadron-hadron and nucleus-nucleus collisions (HIJING, NEXUS,
and PSM). The  HIJING and PSM predictions confirm approximate
energy independence of the entropy dimension $D_1\equiv
S/Y_m=0.417\pm 0.009$ \cite{SSZ} in the full phase-space up to the
LHC energy and even beyond. These models give $S_{LHC}\sim 4.65$
at $\sqrt{s}$=14~TeV and a power-like energy behaviour of the
average multiplicity.
Simulations with NEXUS show monotonous decrease of $S/Y_m$
which corresponds to a slower (logarithmic) increase of the
average multiplicity with energy. In this view,  HIJING and PSM,
unlike to NEXUS, prefer self-similar character of multi-hadron production.
All three models predict approximately constant value of the multifractal
specific heat and thus suggest multifractal character of proton-proton interactions
in the considered energy region.
The information about self-similarity and multifractal structure
is contained in MDs and can be quantified
by the behaviour of both the {\it information} entropy and the
{\it multifractal} specific heat in a model independent way.

\begin{center}
\bf Acknowledgments
\end{center}

This work has been partially supported by the Grant No. ME 475 of the Czech
Ministry of Education, Youth, and Physical Training.

\newpage
\begin{center}
\begin{figure}
\epsfig{file=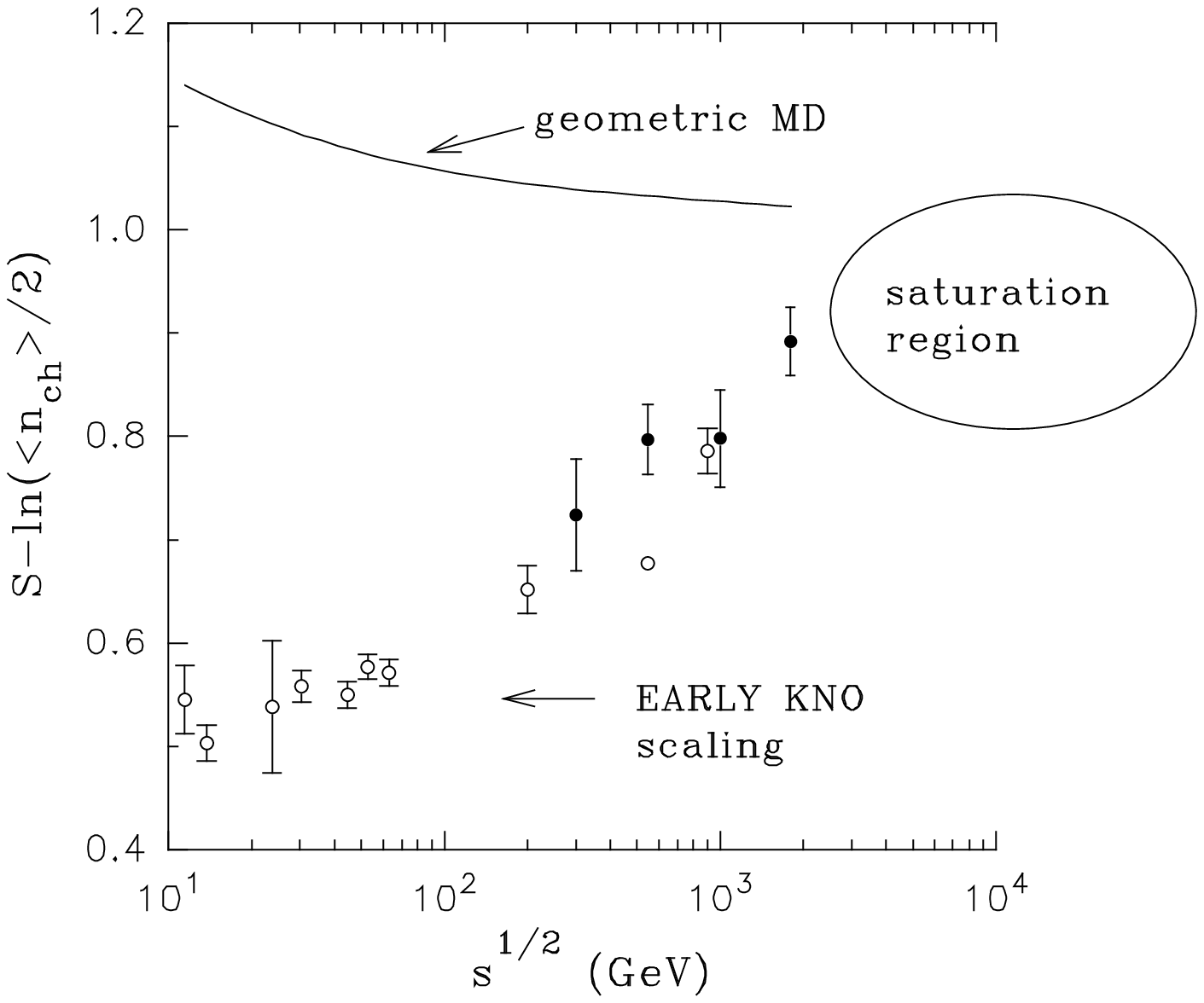, width=90mm}
\caption{
The energy dependence of the difference $S-\ln(<\!n_{ch}\!>\!/2)$
for charged particles (experimental data).
The full circles correspond to data
from E735 experiment.
}
\vspace*{5mm}
\end{figure}
\end{center}

\begin{center}
\begin{figure}
\epsfig{file=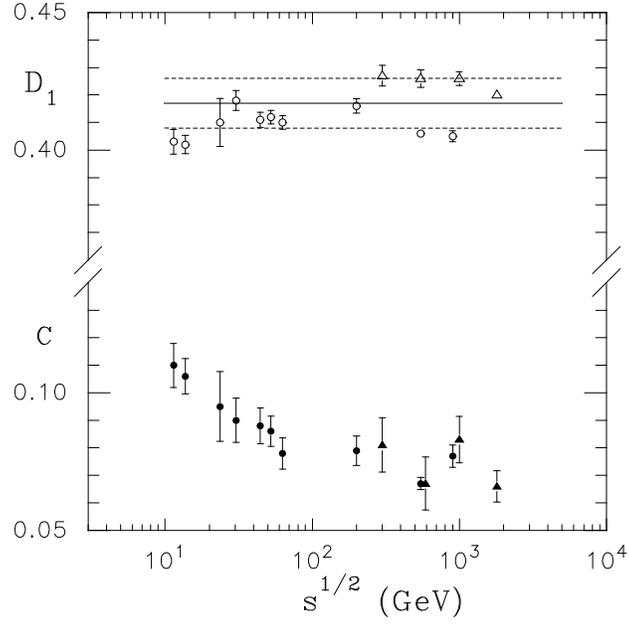, width=82mm}
\caption{
The energy dependence of the entropy dimension $D_1$ and the multifractal specific heat
$c$ for charged particles.
The triangles correspond to data form E735
experiment.
}
\end{figure}
\end{center}

\begin{center}
\begin{figure}
\epsfig{file=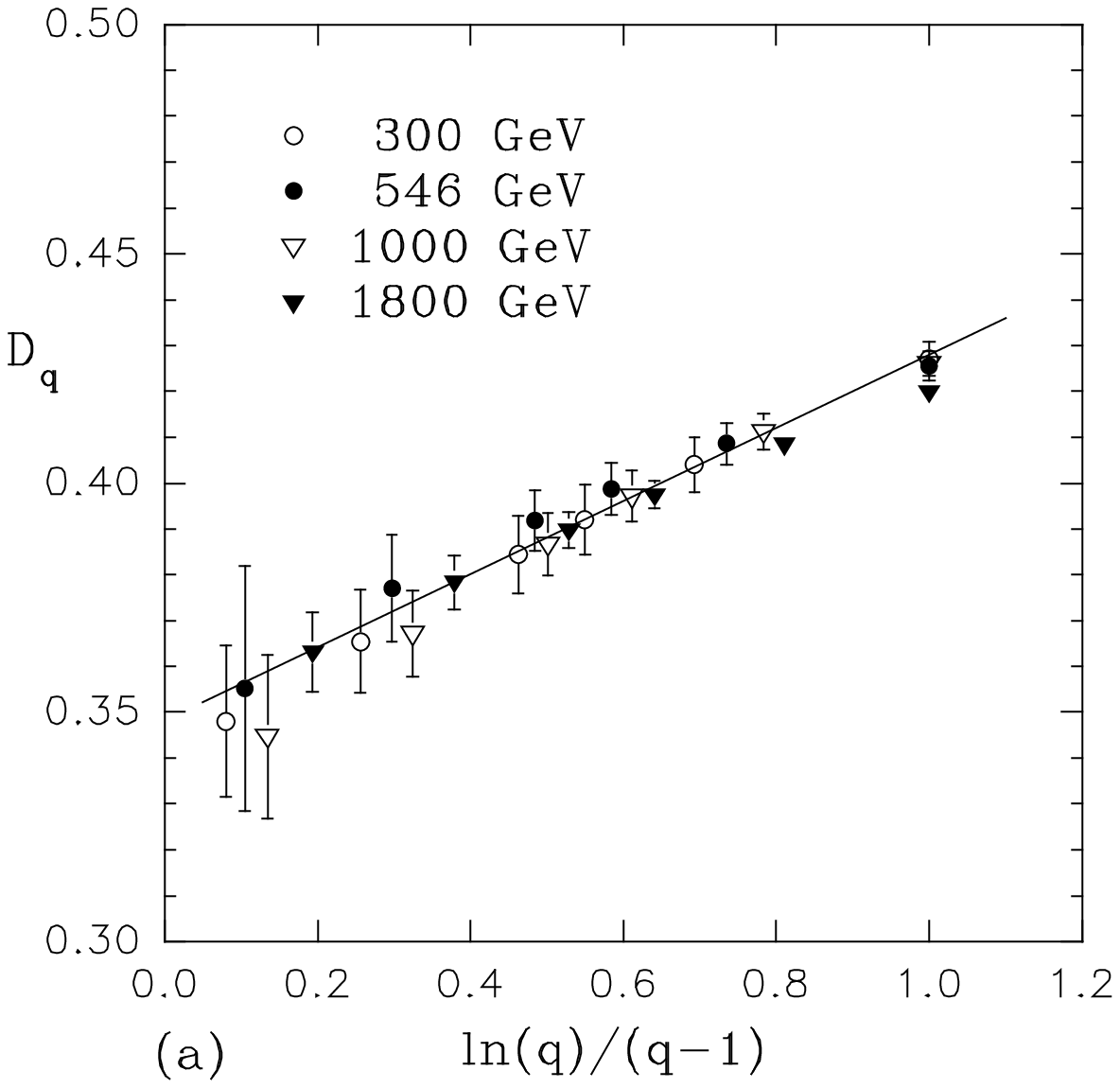, width=90mm}
\end{figure}
\end{center}

\begin{center}
\begin{figure}
\epsfig{file=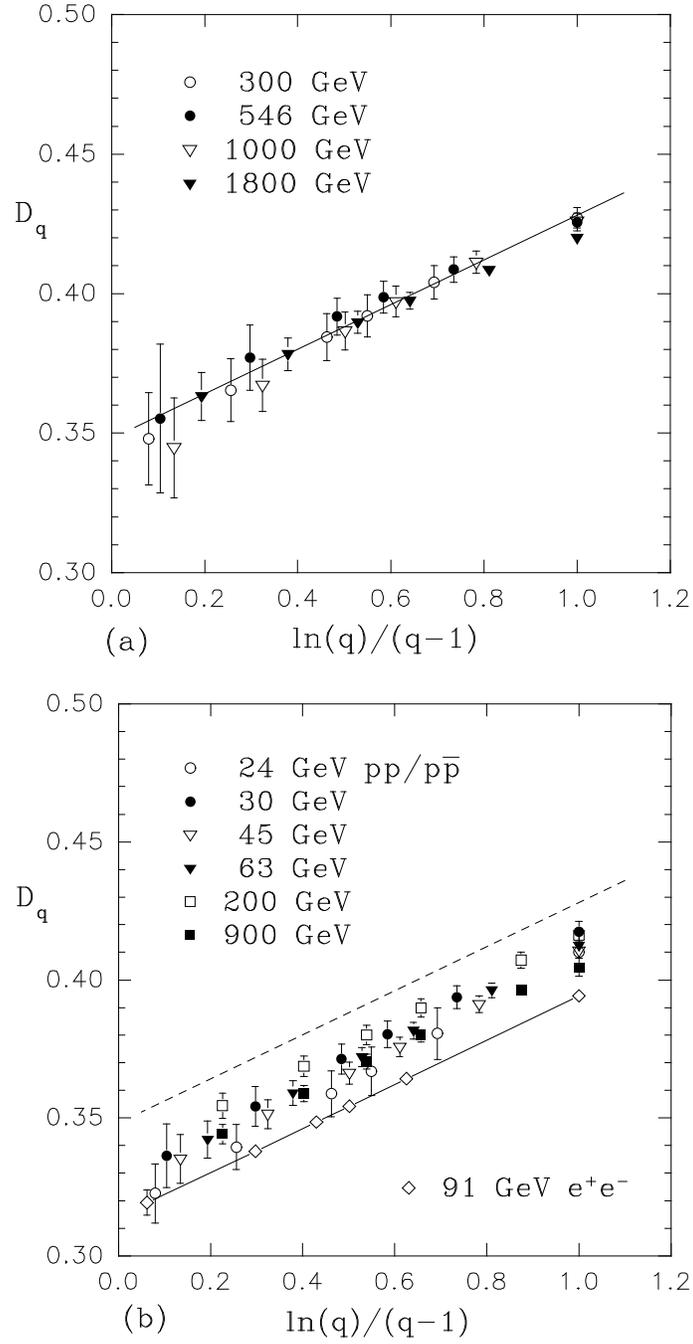, width=90mm}
\caption{
The generalized dimensions $D_q$ as function of ln(q)/(q-1) for
charged particles. Data are taken
(a) from E735 experiment \cite{E735} and (b) from Refs.
\cite{ISR,UA5a,UA5b,LEP}.
}
\end{figure}
\end{center}

\begin{center}
\begin{figure}
\epsfig{file=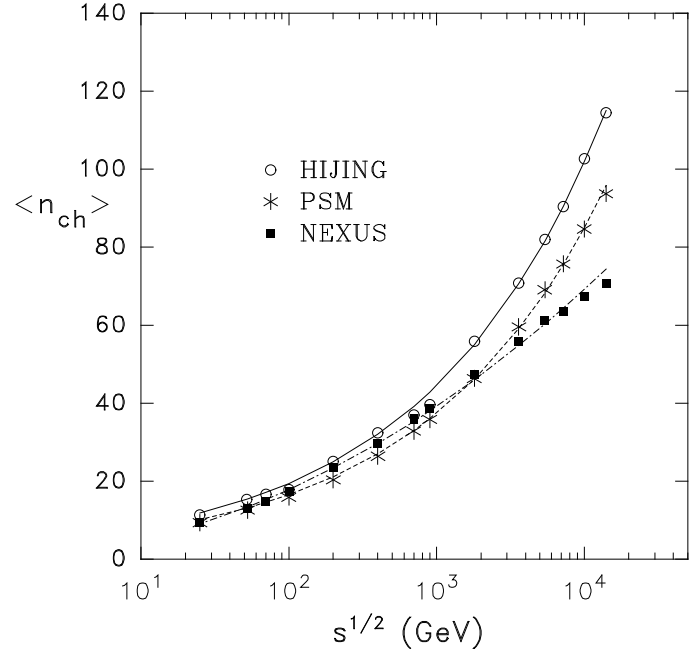, width=90mm}
\caption{
The energy dependence of the average multiplicity simulated by MC for charged particles.
}
\vspace*{2mm}
\end{figure}
\end{center}

\begin{center}
\begin{figure}
\epsfig{file=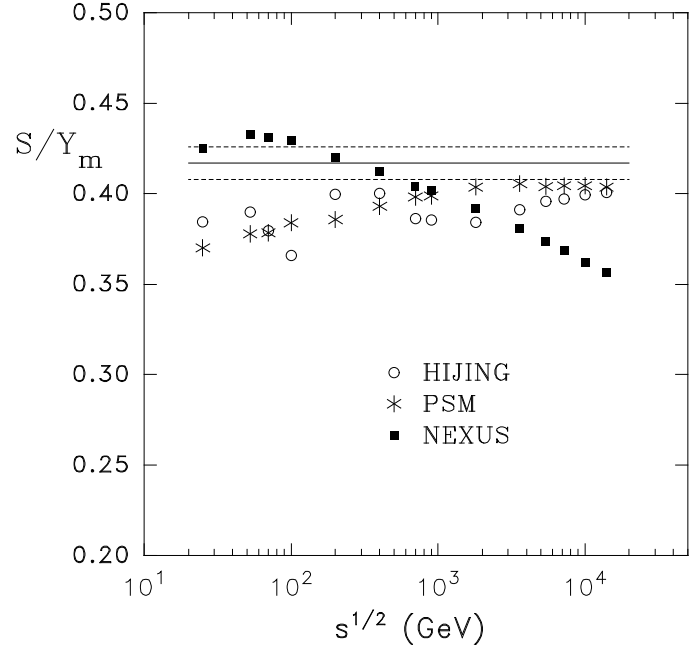, width=90mm}
\caption{
The energy dependence of $D_1=S/Y_m$ of charged particles. The points
correspond to the MC simulations.
}
\end{figure}
\end{center}

\begin{center}
\begin{figure}
\epsfig{file=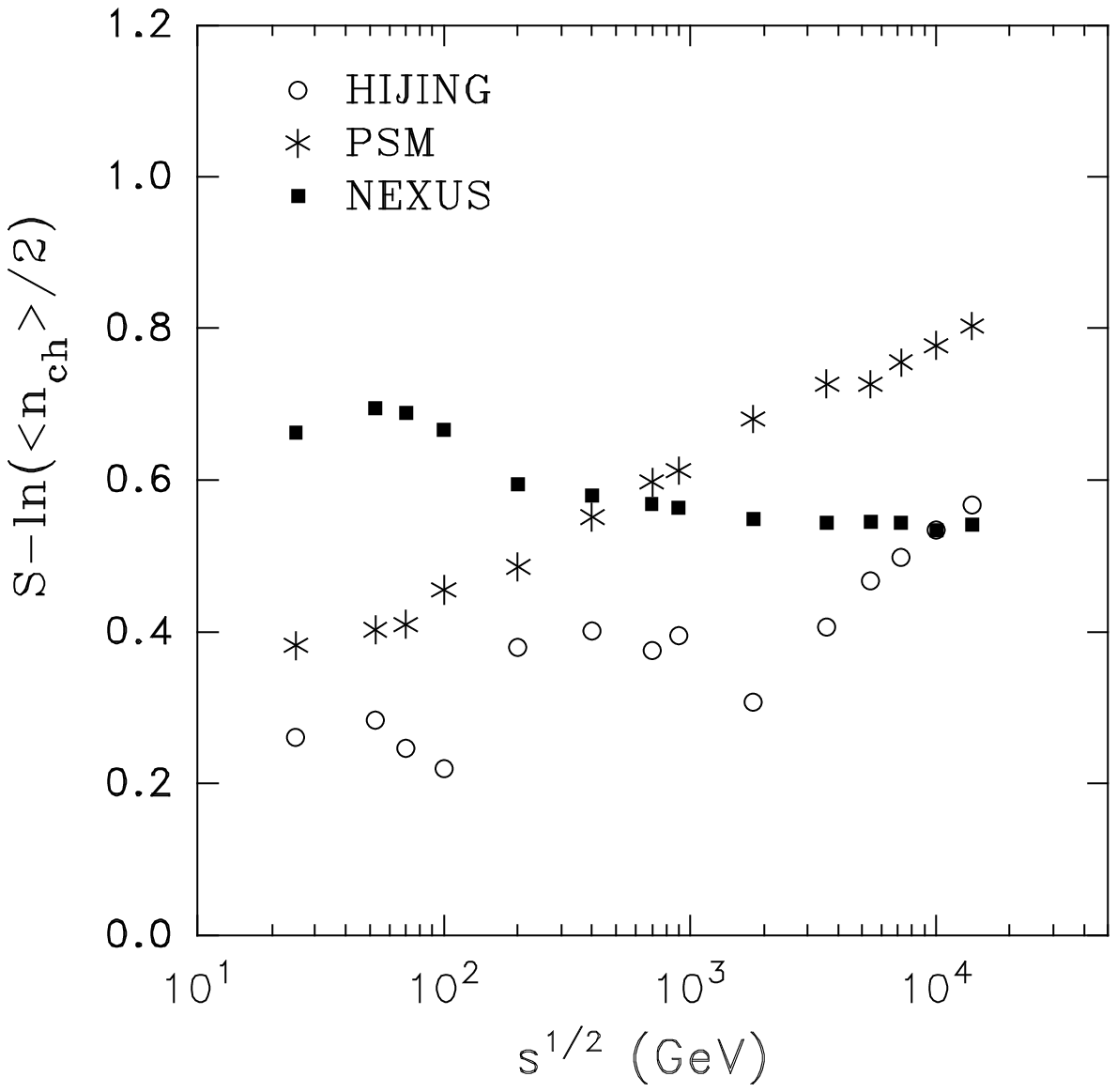, width=90mm}
\caption{
The energy dependence of the difference $S-\ln(<n_{ch}>\!/2)$ of charged
particles. The points correspond to the MC simulations.
}
\end{figure}
\end{center}

\begin{center}
\begin{figure}
\epsfig{file=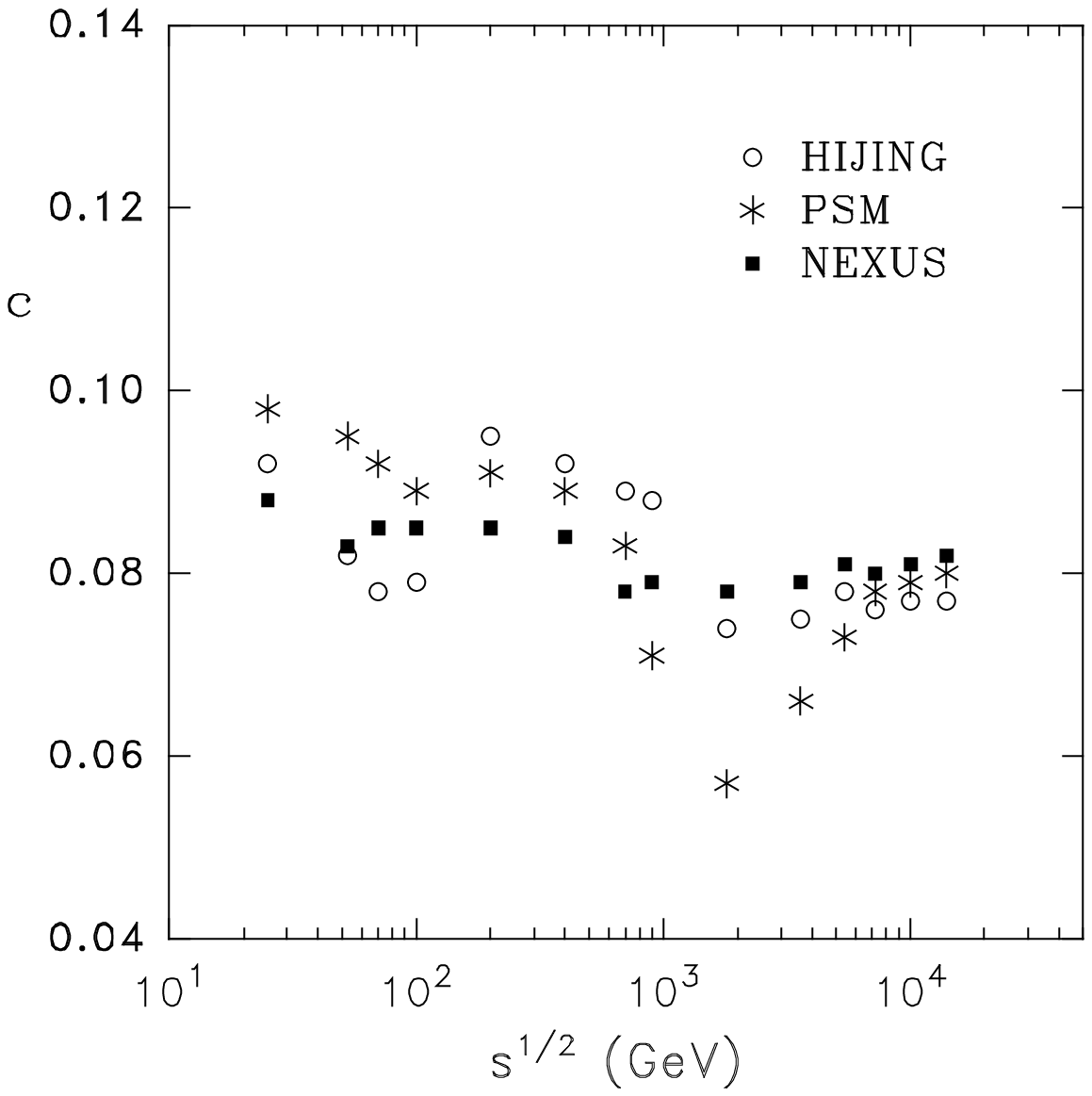, width=90mm}
\caption{
The multifractal specific heat $c$ calculated from MC simulations.
}
\end{figure}
\end{center}


\begin{thebibliography}{99}

\bibitem{Dremin_PR} I.M. Dremin, J.W. Gary, Phys. Rep. {\bf 349}, 301 (2001).

\bibitem{GU99}A. Giovannini, R. Ugoccioni, Phys.Rev. D {\bf 59}, 094020 (1999).

\bibitem{Hwa1}
R.C. Hwa,
Int. J. Mod. Phys. {\bf A4}, 481 (1989); Phys. Rev. D {\bf 41}, 1456 (1990).

\bibitem{Sarcevic}
I. Sarcevic and H. Satz,
Phys. Lett. {\bf B233}, 251 (1989).

\bibitem{Miyamura}
O. Miyamura and T. Tabuki, Z. Phys. C {\bf 31}, 71 (1986).

\bibitem{Bialas}
A. Bialas and R. Peschanski, Nucl. Phys. B {\bf273}, 703 (1986).

\bibitem{Peshanski}
R. Peshanski, Mod. Phys. Lett. A {\bf 6}, 3681 (1991).

\bibitem{Hwa2}
R.C. Hwa and M.T. Nazirov, Phys. Rev. Lett. {\bf 69}, 741 (1992).

\bibitem{DeWolf}
E.A. De Wolf and I.M. Dremin {\it et al.,} Phys. Rep. {\bf 270}, 1 (1996).

\bibitem{Bershadskii}
A. Bershadskii, Physica A {\bf 253}, 23 (1998).

\bibitem{Parashar}
N. Parashar, LI Nuovo Cimento A {\bf 108}, 489 (1995).

\bibitem{Ghosh}
D. Ghosh {\it et al.,} Phys. Rev. C {\bf 65}, 067902 (2002).

\bibitem{Takagi}
F. Takagi, Phys. Rev. Lett. {\bf 72}, 32 (1994).

\bibitem{SSZ} V. \v{S}im\'{a}k, M. \v{S}umbera, and I. Zborovsk\'{y},
Phys. Lett. B {\bf 206}, 159 (1988).

\bibitem{PSSZ} M. Pachr, V. \v{S}im\'{a}k, M. \v{S}umbera, and
I.Zborovsk\'{y}, Mod. Phys. Lett. A{\bf 7}, 2333 (1992).

\bibitem{Renyi} A. R\'enyi, {\it Probability Theory} (North Holland, Amsterdam, 1970);
 A. R\'enyi, {\it Selected Papers of Afred R\'enyi, Vol. 2}
(Akad\'emia Kiad\'o, Budapest, 1976).

\bibitem{Feder} J. Feder, {\it Fractals} (Plenum Press,New York and London 1988).

\bibitem{HIJING}http://www-nsdth.lbl.gov/$\sim$xnwang/hijing/; X.N. Wang
and M. Gyulassy, Phys. Rev. D {\bf 44}, 3501 (1991); Comput. Phys.
Commun. {\bf 83}, 307 (1994).

\bibitem{NEXUS}
http://www-subatech.in2p3.fr/$\sim$theo/nexus/; H.J.~Drescher, M.
Hladik, S. Ostapchenko, T. Pierog, and K. Werner, Phys. Rep. {\bf
350}, 93 (2001); Phys. Rev. Lett. {\bf 86}, 3506 (2001).

\bibitem{Werner_QM01}K. Werner, H.J. Drescher, S. Ostapchenko,
and T. Pierog. Nucl. Phys. A {\bf 698}, 387 (2002).

\bibitem{PSM} N.S. Amelin, N. Armesto, C. Pajares, D. Sousa, Eur.
Phys. J. C {\bf 22}, 149 (2001).

\bibitem{KNO} Z. Koba, H.B. Nielsen, and P. Olesen,
Nucl. Phys. B {\bf 40}, 317 (1972).

\bibitem{ISR} A. Breakstone {\it et. al.,} Phys. Rev. D {\bf 30}, 528 (1984).

\bibitem{UA5a} R.E. Ansorge {\it et. al.,} UA5 Collaboration,
Z. Phys. C {\bf 43} {\it Particles and Fields}, 357  (1989).

\bibitem{UA5b} UA5 collab., Phys. Rep. {\bf 154}, 247 (1987).

\bibitem{E735}  T. Alexopoulos {\it et al.,}  Phys. Lett. B {\bf 435}, 453 (1998).

\bibitem{Pietronero}F. Sylos Labini, M. Montuori and L.Pietronero,
Phys.Rep. {\bf 293}, 61 (1998).

\bibitem{Paladin} G. Paladin and A. Vulpiani, Phys.Rep. {\bf 156}, 147 (1987).

\bibitem{Landau}
L.D. Landau, E.M. Lifshitz, {\it Statistical Physics, Part 1}
(Pergamon Press, Oxford, 1980).

\bibitem{LEP}
DELPHI collab., Z. Phys. C {\bf 52}{\it Particles and Fields}, 271
(1991).

\bibitem{Bautin}
A.V. Bautin, Physics-Uspekhi {\bf 38}, 609 (1995) (English
translation, AIP, New York).

\end{thebibliography}
\end{document}